\documentclass[%
 reprint,
 amsmath,amssymb,
 aps,
prb,
]{revtex4-1}

\usepackage{graphicx}
\usepackage{dcolumn}
\usepackage{bm}

\begin{document}

\preprint{APS/123-QED}

\title{A Model for Correlated Metamagnets: Application to  UPt$_3$ and CeRu$_2$Si$_2$}
\thanks{Dedicated to my Parents}%

\author{Bellave S. Shivaram}

\affiliation{ Department of Physics, University of Virginia, Charlottesville, VA. 22901}

\date{\today}

\begin{abstract}

In a model system with three S=1 pseudo-spins situated in an anisotropic mean field we compute the thermodynamic response in a self consistent manner. An adhoc uniform broadening of all energy levels of the (localized) spins due to their interaction with conduction electrons is introduced. The results with a minimal set of parameters reproduce with remarkable fidelity the properties of the strongly correlated metamagnets UPt$_3$ and CeRu$_2$Si$_2$. Several key predictions are made for further experimental confirmation.

\begin{description}

\item[PACS numbers]

75.30.Mb, 75.20.Hr

\end{description}

\end{abstract}

\pacs{Valid PACS appear here}
\maketitle

Strong electronic correlations are a hallmark of many different types of fermionic systems. Electrons in many metals and semimetals belonging to the heavy fermion family, the high T$_c$ cuprates, pnictides, organic charge transfer salts and fermions in liquid $^3$He are some examples \cite{BalentsRev2014, GeorgesRev2012, KatoRev2011, FuldeRev2006}. While much progress has been made in recent years a full quantum mechanical treatment of the many body problem is still a challenge. On the other hand phenomenological models that capture the core common signatures in these varied materials can be tremendously helpful in facilitating further experimental discoveries. The occurance of quantum phase transitions in the classes of materials mentioned above is one such common signature \cite{SachdevQPT, QiSteglichScience2010, BoebingerQPTPRL2009, LohneysenWolfleRMP2007}. Here, the transitions driven by quantum fluctuations are sensitively tuned by an external parameter such as pressure or magnetic field. Systems with antiferromagnetic (AFM) fluctuations rather than ferromagnetic (FM) seem to be preponderant \cite{AndersonAdvPhys1997} in many of the recently discovered classes of materials .  The presence of a strong anisotropy in the magnetic properties in many of the systems is another common theme \cite{FradkinNematicRev2010}. \\

Any metal without exchange/correlations between spins exists in a non-ordered state normally referred to as paramagnetic. In a paramagnet in the absence of thermal disturbance an infinitisemally small field can align all spins and the linear susceptibility diverges at T = 0. In a strongly correlated electronic system on the other hand quantum fluctuations can result in a finite susceptibility at absolute zero. Concurrently, such fluctuations can prevent the condensation of true long range magnetic order and the spins can attain a 'metamagnetic state'.  Several members of the families of metals mentioned above such as UPt$_3$ and CeRu$_2$Si$_2$ in the case of heavy fermion materials \cite{FranseFringsPhysicaB1984}, LaCu$_2$O$_4$ \cite{ThioLa2CuO41988}, \cite{CheongFiskLa2CuO41989} among the cuprates, other oxides \cite{PerryPhysRevLett2001, BingSigristEPL2001, NakatsujiJLTP1999}, and several pnictides and chalcogenides \cite{BaltzerCrSpinel1966,ImaiSrCaCoP2012, AnandCaCoAsPRB2014, YingCaSrCoAsPRB2012} maybe classified as metamagnets. In these systems the magnetic response which starts out linearly in small fields is found to rise sharply in a nonlinear fashion at a critical magnetic field.  In recent work a remarkable degree of success has been achieved in describing the behavior of such correlated metamagnets \cite{ShivaramChi3, ShivaramChi5, 
ShivaramHighFieldUS} with a model involving a single energy scale (SES) \cite{KumarCelliShivaram2015}. In the SES model there is an excited magnetic doublet (or a triplet) which splits in a field. The lower of the split states evolves to cross the non-magnetic ground state at a critical field, H$_c$, resulting in a sharp rise in the magnetization, a characteristic signature of metamagnetism. The gap between the excited magnetic state and the ground state sets the single energy scale, $\Delta$, and also gives rise to other characteristic signatures. For instance, there is a peak in the linear susceptibility at a temperature T$_1$ which scales with H$_c$ and also a peak in the third order susceptibility at a temperature T$_3$=0.4T$_1$ all of which are faithfully reproduced in experiments \cite{ShivaramChi3}. However, all magnetic susceptibilities produced by the SES model tend to zero as T $ \rightarrow$ 0. This is in sharp contrast in UPt$_3$ for example where the zero temperature susceptibility drops only to $\approx$ 80\% of its value\cite{FranseJMMM1985} at T$_1$. In addition, it is known that many correlated electronic systems exhibit typically a large negative Curie-Weiss temperature, $\theta_{CW}$, determined from the intercept of the high temperature inverse susceptibility. This may be interpreted as the result of strong AFM correlations that start out at high temperatures and are the underlying basis of low temperature order including superconductivity that eventually evolves. The SES model does obtain a negative intercept when $\chi_1^{-1}$ = 0 but the magnitude predicted, $\Delta/3$, is too small compared to typical values $\approx 5 \Delta$ observed in several metamagnets.  Forced to examine these outstanding issues we have arrived at a closely related model which is able to address these aspects as well as many other important experimental observations. In this paper we present the results from this model and demonstrate its effectiveness by accounting for the thermodynamic response of the strongly correlated electronic systems, UPt$_3$ and CeRu$_2$Si$_2$. The model with trivial modifications should apply equally well to other correlated electronic systems. \cite{ShivaramSMM, ShivaramRuthenate}.  \\

We begin by considering three non-interacting pseudo spins (S=1) at a single site to which a magnetic field is applied.  The Hamiltonian of this system can be written as:\\

 $ H=\sum\limits_{i=1,2,3}( \Delta_i  S_{iz}^2 - g_i h S_{iz})$       \hspace{8em}           (1)\\

Here the g$_i$ are proportional to the magneto-gyric ratio and $\Delta_i$ at the very least are a measure of the anisotropy of the pseudo spins S$_i$. The magnetic field h in the present work is considered separately to be either parallel or perpendicular to the quantization axis (the z axis). Clearly, with the inclusion of only the z-component in the Hamiltonian the model favors anisotropic materials or systems which are predominantly not cubic. Although not obvious at this time as we will demonstrate later this model Hamiltonian begets a large negative $\theta_{CW}$ in a convenient manner. It also produces a large zero temperature susceptibility simply because of the possibility that one of the $\Delta_i$ can be chosen to be small. \\

In fig.1 we show schematically the energy eigenvalues obtained from (1) when the magnetic field is in the z- direction. There is a low energy scale $\Delta_1$ and an intermediate scale set by $\Delta_2$. A comparison of these energies with the $\Delta$ in the SES model is appropriate at this point. As mentioned above, in the SES model there is a jump in the magnetization at a critical field which scales with $\Delta$, and there is a peak in the linear susceptibility at a temperature \cite{Note1} $t_1 = (2/3) \Delta$.  In general, the introduction of additional terms (spins) in the Hamiltonian will append to the number of crossing levels and potentially introduce new features in the magnetization isotherm. Concomitant alterations in the susceptibilities can also be expected.  However, on the low temperature/low field side any new features such as additional peaks or steps can be "washed" out by introducing broadened energy levels. We do so in our model with a parameter, 'w', comparable in magnitude to $\Delta_1$.  The physical justification for introducing level broadening comes from experiments \cite{Otte2008}.   All the features established at the energy scale $\Delta_2$ one hopes survive more or less intact but with appropriate modifications coming from the 'proximity' of $\Delta_1$. The third energy scale $\Delta_3$ is chosen such that $\Delta_3 >> \Delta_2$ also to preserve all features arising from the intermediate energy scale, $\Delta_2$. The key point here is that although there are several energy scales the many body interactions (via the level broadening) ensure that effectively only a single energy matters. Thus, for example, external perturbations such as pressure or chemical composition while altering the response do so in such a way that all observables at low temperature scale on a single parameter\cite{Note1, FuldeThalmeier1986, AuerbachLevin1986, PuechJLTP1988, ContinentinoScaling1991} . Of course there could be variations in behavior that depend on the individual values of all the energies, $\Delta_1, \Delta_2, \Delta_3$ and w but if the relative values can be kept more or less within a reasonable range then one can expect the same general trends in all real systems. Furthermore, we require that external perturbations do not disturb the essential Ising nature of the environment and the predominanty AFM nature of the on-site correlations. However,  while the former is a strict necessity the latter condition can be relaxed, allowing for the possibility of FM interactions between the bare spins. Thus truly the only requirement is an Ising character of the environment in which the spins find themselves.\\

 \begin{figure}
 \includegraphics[width=90mm]{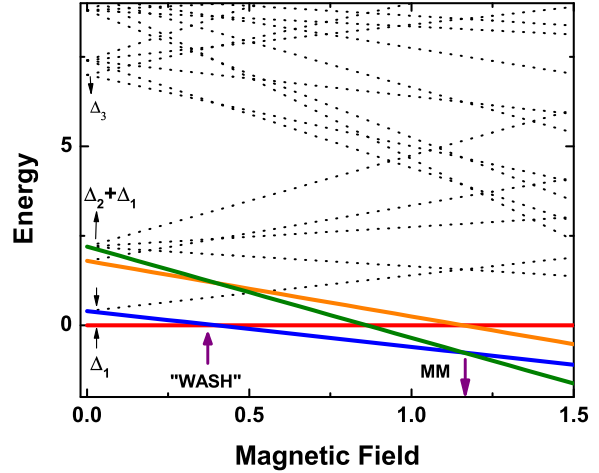}
\caption{\label{fig1} Shows the energy level scheme obtained from the Hamiltonian (1) when the field is applied parallel to the quantization axis ( $h=h_z$ ).  There is a non-magnetic ground state (red) at zero, followed by a first excited magnetic doublet at  $\Delta_1$ (blue).  A third set of magnetic levels cluster at $\Delta_2$  and at $\Delta_1+ \Delta_2$. There are also additional energy levels clustered around $\Delta_3$. Although there are several level crossings the first state to cross the nonmagnetic ground state in general produces metamagnetic features.  However, the first crossing labeled ''WASH' is washed out and the next one at the higher field labeled MM results in the nonlinear rise of the magnetization}.
\end{figure}

In the specific model chosen in (1), which for ease of reference we will term as the SES+ model, there are several level crossings and in general a jump in the magnetization is expected whenever a magnetic excited state crosses the non-magnetic ground state. But with many level crossings what counts is the first level to do so as the field is ramped. Thus in fig.1 the arrows indicate those level crossings that could in principle be responsible for a jump in the magnetization. However, only the one labelled 'MM' contributes to a nonlinear rise in the magnetization at the corresponding critical field.  \\

The three S=1 spins possess twenty seven energy levels and these enter the partition function Z. In order to represent the interaction of the localized spins with the conduction electrons we introduce a hybridization parameter 'w' the effect of which is to replace the temperature t by $t_w =\sqrt{t^2 + w^2}$ in the Boltzmann factor as well as in the expression for the free energy F = - $k_Bt_wln(Z)$ \cite{NagaokaCrommiePRL2002}. The magnetization m, is calculated from the free energy as $m =- \partial{F}/\partial{h}$ where 'h' is the model magnetic field and with the standard procedure of replacing the field h by h + $\lambda$ m, where $\lambda$ is the mean field parameter, one can obtain its self consistent values. The effect of the hybridization parameter, w, is to 'wash out' the first level crossing and only the second crossing marked 'MM' in fig.1 results in a magnetization jump.  In addition to the magnetization other thermodynamic quantities can be evaluated as well, and we demonstrate this by considering the heat capacity further below in this work.\\

The following is the ordering of the paper: the linear and nonlinear magnetic response for UPt$_3$ is presented for the parallel case, h$\parallel $z-axis, followed by a similar treatment for the perpendicular geometry, h$\parallel$x - axis. A discussion of the high temperature response for both geometries in UPt$_3$ as well as CeRu$_2$Si$_2$ is presented next. This is followed with results for the heat capacity and a treatment of the pressure and composition dependence of the magnetic properties of both UPt$_3$ and CeRu$_2$Si$_2$. Finally a discussion of all results is presented with suggestions for further experiments.\\

\begin{figure}
\includegraphics[width=90mm]{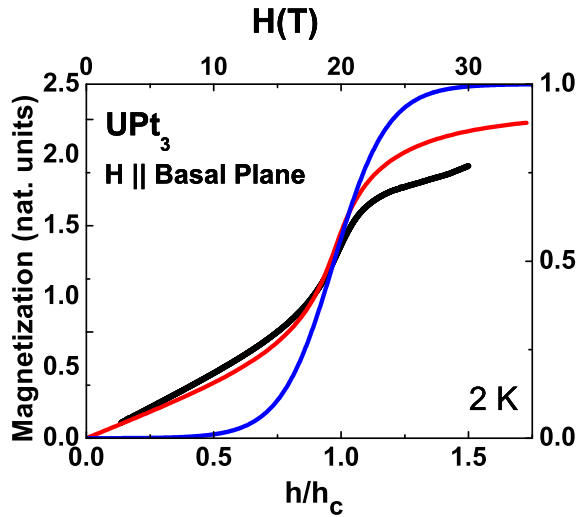}
\caption{\label{fig2} Magnetization isotherm for T=2 K. The black line is the experimental data, the red line is from the present model with model parameters $\Delta_1 = 0.40, \Delta2 = 1.8, \Delta_3 = 8, w = 0.34, g_1 = 0.8,  g_2 = 1.5 and g_3 = 2$. The blue line is from the SES model. The initial nlinear rise in the magnetization with a large slope preceeding the metamagnetic transition is clearly reproduced in the present model.}
\end{figure}

\textbf {Magnetic Response - Parallel Case:}  In fig. 2 we show the measured magnetization in UPt$_3$ along with the model results evaluated by choosing the parameters \cite{Note1} $g_1 = 0.8, g_2 = 1.5, g_3 = 2, \Delta_1 = 0.4, \Delta_2 = 1.8$, and $\Delta_3 = 8$ in the Hamiltonian (1) and w=0.34. To create this plot the model results are scaled in two ways - once along the horizontal axis so that the experimental critical field, H$_c$, matches that found in the model, h$_c$, and secondly once along the vertical axis so that the initial rise in the magnetization (the linear susceptibility) at a specific temperature matches again the experimental result. In addition a calibration of the model temperature also needs to be carried out.  This is done by matching the temperature of the peak in the linear susceptitibility at $t_1$ to the measured temperature $T_1$.  In fig. 2 the blue line is from the SES model and the red line is the current model calculated with 'natural units' $\mu_B = 1$ and k$_B$ = 1. In obtaining this response we found that it is necessary to include a ferromagnetic mean field which enhances the response in the vicinity of the metamagnetic transition. The necessity for this mean field is further made clear in fig. 3 where we show the differential susceptibility calculated from the model. Also shown in this figure are the results without the addition of the mean field. In the latter case the response is very broad and we are not able to match the observed magnetic behavior at different temperatures. On the other hand the inclusion of a mean field, $\lambda$, which alters the perceived magnetic field to h + $\lambda$m, with a value $\lambda$ = 0.4 is able to better fit the experimental results.\\

\begin{figure}
 \includegraphics[width=70mm]{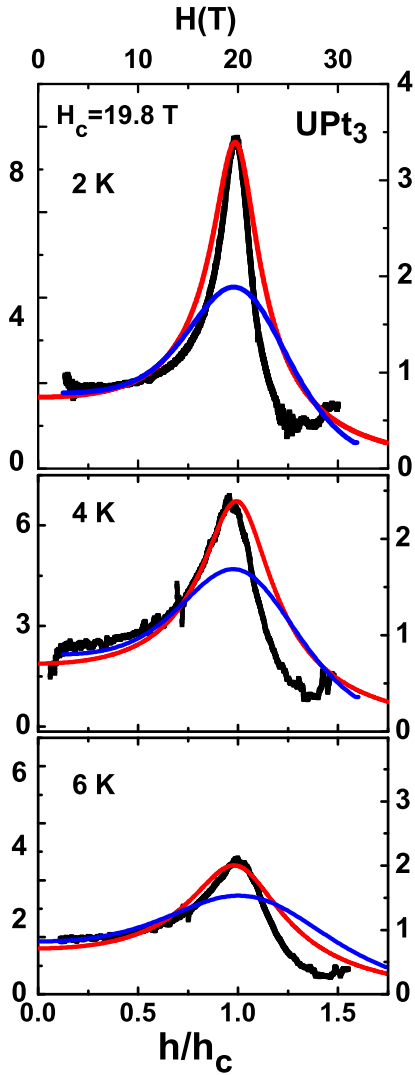}
\caption{\label{fig3}
Differential susceptibility obtained from the measured magnetization isotherms compared to the model results. The ''bare'' model results, shown by the blue lines for all three temperatures do not yield a sharp enough response at the metamagnetic transition. To account for the observed sharpness a ferromagnetic mean field is introduced for this direction. With a mean field parameter $\lambda = 0.4$ the critical field h$_c$ in natural units shifts to a lower value of 0.62. This value sets the field calibration and corresponds to H$_c$ = 19.7 T observed in the experiments.}
\end{figure}

\begin{figure}
 \includegraphics[width=90mm]{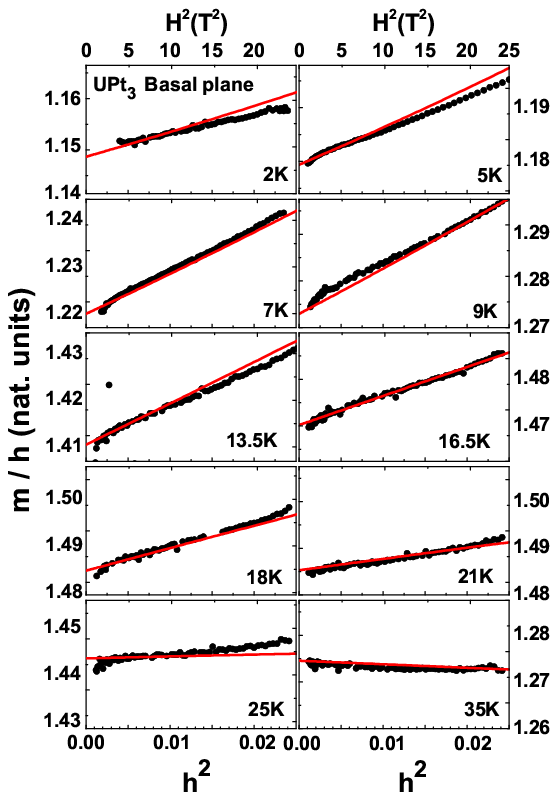}
\includegraphics[width=85mm]{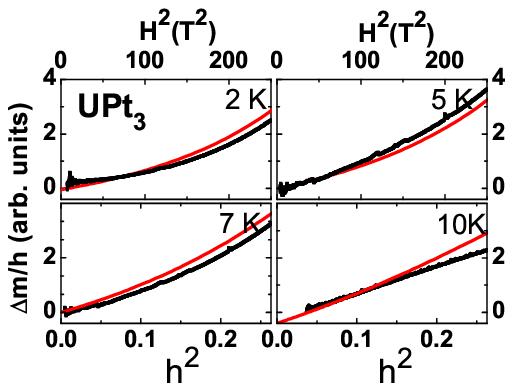}
\caption{\label{fig4}
Illustrates the procedure employed to extract the non-linear part of the response in UPt$_3$. The top panel demonstrates (field upto 5T only) that the slope which is a measure of $\chi_3$ at 2K starts with a lower positive value, increases as the temperature is raised to reach a peak and finally at higher temperatures is negative. The H=0 intercept in these plots yields the linear susceptibility, $\chi_1$. The lower panel is similar but with the magnetic field extending to 16 T so that the contribution from the next higher susceptibility, $\chi_5$ is apparent. Red lines are from the model and the black dots are the experimental points. }
\end{figure}

In addition to sharpening the transition the effect of the mean field is to shift the metamagnetic (MM) rise from occuring at the nominal value of h=1 in natural units to lower values of h for larger $\lambda$. The position in h where the MM transition occurs (defined as the peak in the differential susceptibility) is a convenient way to identify the "mile marker" to calibrate the natural units in terms of the experimental units.  This calibration is fixed for a given sample but will change when parameters such as pressure or composition are altered. We thus have potentially eight parameters, the three $\Delta$'s, the three g-factors, $\lambda$ and 'w', that can be varied. But sets of these once chosen are held constant as we consider changes in orientation, pressure and composition. The three energy levels for instance cannot change as we consider the magnetic properties with rotation of the field. The g-values are also held fixed with rotation and in addition remain unaltered when pressure is applied. Thus, as will be seen repeatedly even though there are many parameters we only need to adjust a few of them to account for all observed properties and the single energy scale behavior is largely preserved.\\

\begin{figure}
 \includegraphics[width=95mm]{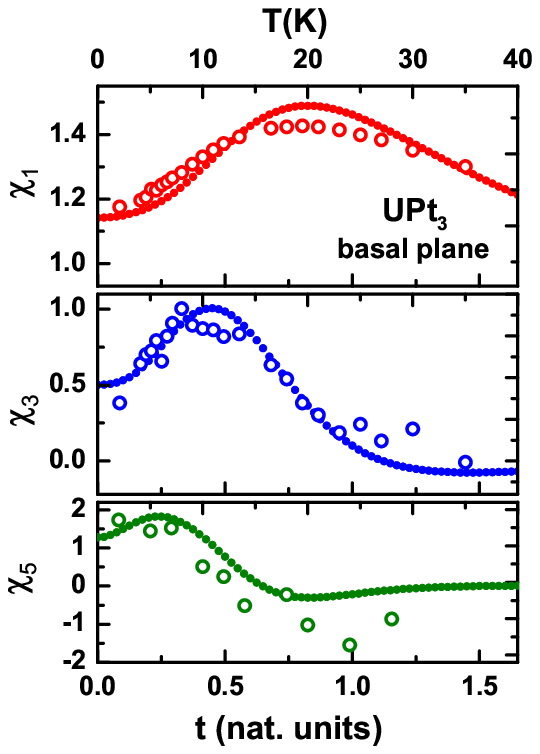}
\caption{\label{fig5}
Shows all three susceptibilities $\chi_1$ (top panel), $\chi_3$ (middle panel) and $\chi_5$ (bottom panel) obtained from experiments and a comparison with the model results in natural units. It is clear that the model performs extremely well in accounting for all the key experimental signatures. The peak in $\chi_3$ in the model occurs precisely at 0.5 of T$_1$ where there is a peak in $\chi_1$. $\chi_5$ also shows a weak maximum in the model, but the data presented is ambiguos about this feature. It will be shown below that the same parameters also account successfully for the
magnetic behavior in the perpendicular geometry.}
\end{figure}

To parallel the experimental procedure employed to extract the nonlinear and linear magnetic susceptibilities we present in fig. 4 the model m/h as a function of $h^2$. For fields corresponding to less than 5 T for UPt$_3$ such a plot (top part of fig.4) yields the third order susceptibility $ \chi_3$ and going beyond 5 T we can also quantify the fifth order susceptibility, $\chi_5$ as we showed earlier \cite{ShivaramChi5}. The three susceptibilities obtained from the model along with the experimental results on UPt$_3$ as a function of temperature are shown in fig.5. The experimental correlation of $T_3/T_1 = 0.5$ is clearly reproduced in this instance and will be seen to hold good later in CeRu$_2$Si$_2$ also for a variety of parameter values. While the model shows a weak very low temperature peak in $\chi_5$ this feature was not discerned in our experiments \cite{ShivaramChi5}. More precise measurements of $\chi_5$ are required and may be able to pin down the presence or absence of this feature. As pointed out earlier the measured $\chi_5$ taken together with $\chi_3$ and $\chi_1$ implied a possible thermodynamic instability in UPt$_3$. This was ascertained by checking if the stability parameter $(3 \chi_3^2-\chi_5 \chi_1)$ is always greater than zero. We plot this quantity evaluated in the current model in figure 6. The stability parameter is positive at high temperatures as in the SES model but turns negative at t $\approx$0.3 in fair agreement with experiment. All of these results demonstrate the overwhelming success of the present model. \\

\begin{figure}
 \includegraphics[width=90mm]{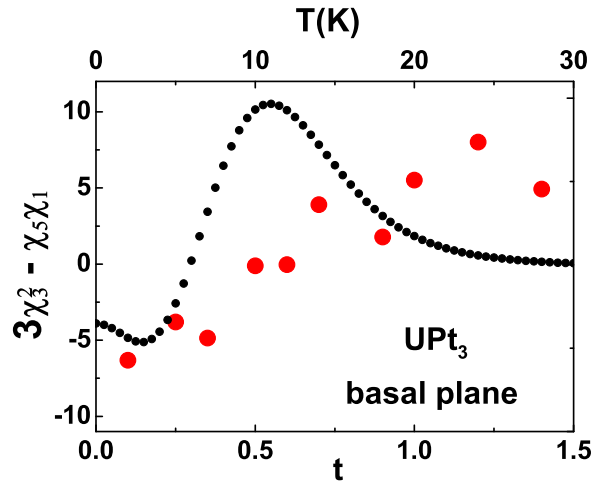}
\caption{\label{fig6}
The measured susceptibilities imply a violation of the stability condition (see text). This is brought out clearly in the present calculations in contrast to the SES model.}
\end{figure}

\begin{figure}
\includegraphics[width=90mm]{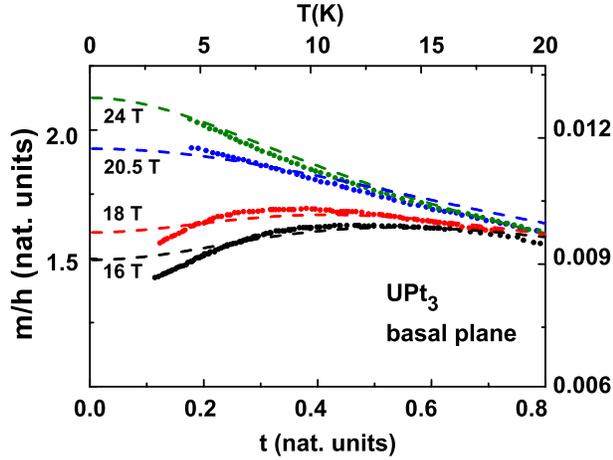}
\caption{\label{fig7}
The measured temperature dependence of the magnetization at high magnetic fields close to the metamagnetic critical field. The magnetization at the critical field is almost temperature independent as $t \rightarrow 0$. For a fixed field on the high side of the critical field the slope is negative and on the low side the slope is positive. These features as well as the shift in the position of the maximum in the susceptibility to lower temperatures as the field is increased are all brought out elegantly in the model. The experimental points are from Kim et al. \cite{KimStewartSSC200}.}
\end{figure}

The above discussion has focused primarily on the low field properties. Turning our attention to the magnetic response in the vicinity of the MM transition we  can examine results by other experimentalists. Kim and Stewart \cite{KimStewartSSC2000} performed temperature dependent magnetization measurements at constant fields upto 24 T. Their results along with model calculations employing the same parameters as above (ref. fig.3) are shown in fig. 7. Here again the calculated trends are in outstanding agreement with the experimental results. The peak in the magnetization which occurs at T$_1$ in low fields moves to lower temperatures, is eventually suppressed and the response at the critical field is flat at the lowest temperatures. \\

\begin{figure}
\includegraphics[width=90mm]{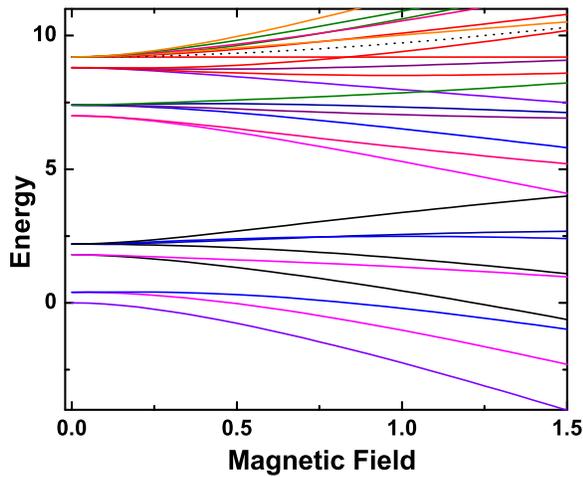}
\caption{\label{fig8}
Illustrates the magnetic field evolution of the energy levels when the field is along the x-axis. There are no level crossings such that a magnetization transfer to a low lying state can occur. Consequently a metamagnetic transition would not be expected for this geometry.}
\end{figure}

\begin{figure}
\includegraphics[width=90mm]{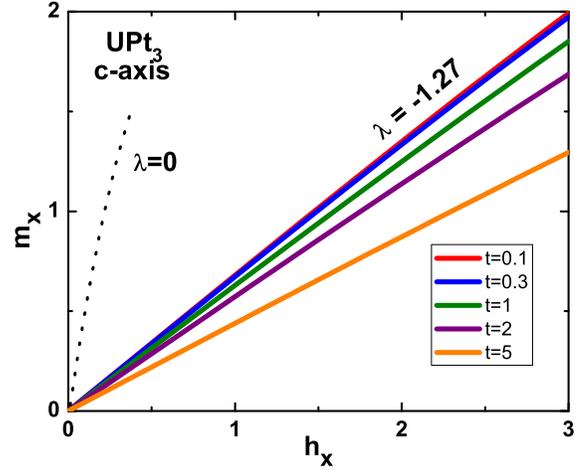}
\caption{\label{fig9}
Illustrates the evolution of the magnetization when the field is along the x-axis for five different temperatures marked in the figure. A metamagnetic transition would not be expected for this geometry and the magnetic response appears deceptively similar to that of a paramagnet. However, there are significant differences when the nonlinear response is considered. Also shown is the response expected when there is no AFM mean field (dotted line). }
\end{figure}

\begin{figure}
\includegraphics[width=95mm]{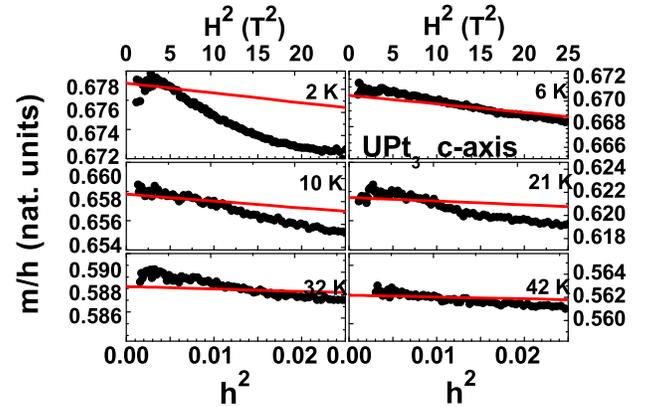}
\caption{\label{fig10}
This figure similar to fig.4 shows the experimental results compared with the model m/h vs $h^2$.  The slope of the lines, a measure of $\chi_3$ is negative at all temperatures.  The model underestimates it by approximately a factor of two at the lower temperatures.}
\end{figure}

\begin{figure}
\includegraphics[width=90mm]{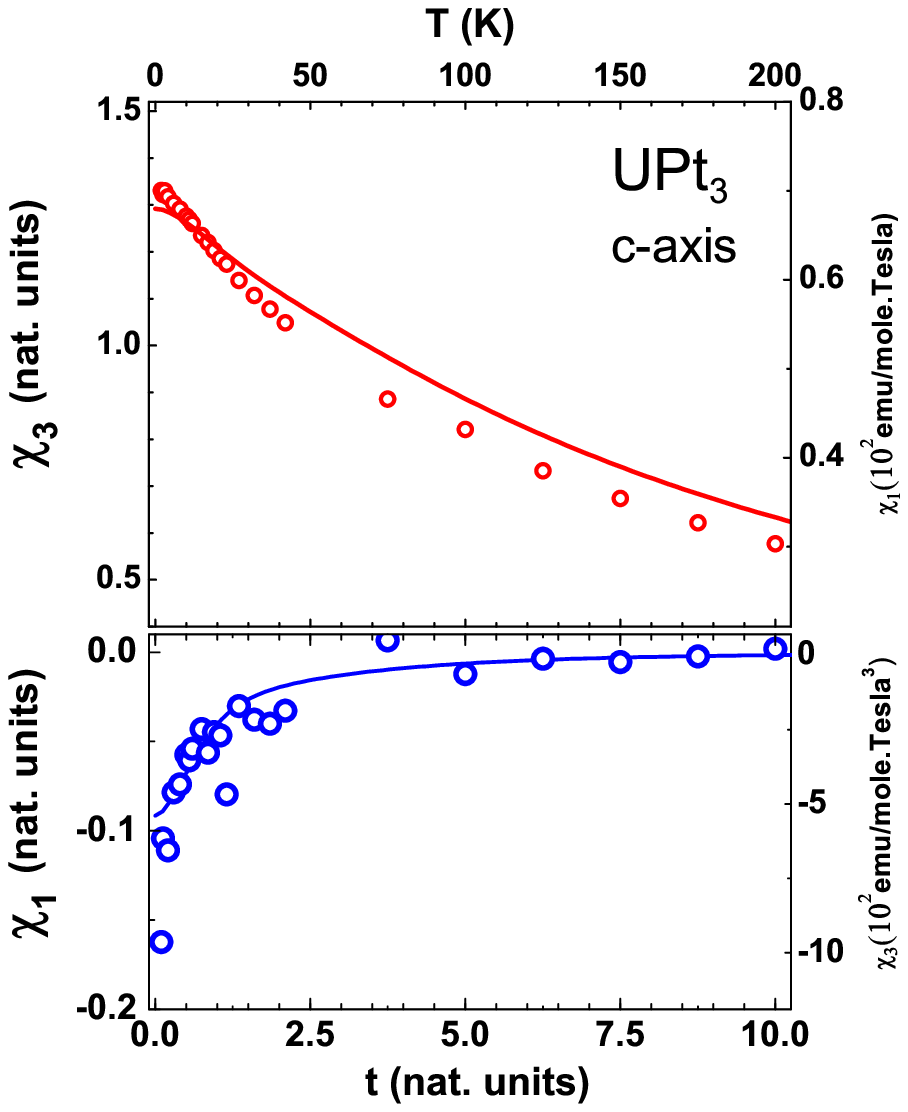}
\caption{\label{fig11}
Shows the evaluated $\chi_1$ and $\chi_3$ in UPt$_3$ as a function of temperature with the same parameters as before used in the parallel case.  However, here a mean field parameter $\lambda=-1.27$ is used to reduce the linear susceptibility at t=0 to the value shown. As a consequence the model $\chi_3$ is also significantly reduced and at the lowest temperatures is $\approx 2$ times smaller than experimental values.}
\end{figure}

\textbf{Magnetic Response - Perpendicular Geometry:} The energy levels evaluated for this geometry are shown schematically in fig.8. The general structure of these energy levels is similar to that in the SES model and therefore we would expect the calculated magnetic response to be similar. In fig.9 we show the magnetization isotherms obtained for this geometry. It should be noted that although the behavior is similar to a paramagnet there are significant differences in how the magnetization evolves in a correlated metamagnet in this orientation. This point becomes clear as we consider the higher order susceptibility, $\chi_3$ apparent in fig. 10 and shown explicitly together with $\chi_1$ in fig.11.  A Curie type behavior in $\chi_1$ at high temperatures gradually evolves at lower temperatures such that the linear susceptibility saturates to a finite value at t=0. This value would be much larger than that obtained at the peak for the parallel geometry if the same positive mean field parameter is employed. This is inconsistent with the experimental observation in UPt$_3$ where $\chi_1(0)$ for H $\parallel$ c-axis is less than half the value at the peak for H $\perp$c-axis \cite{FringsFranseUPt3Chi11983}.  In order to match these values to experiments it is necessary to invoke an antiferromagnetic (AFM) mean field in this direction, fairly significant in magnitude. This value, $\lambda=-1.27$ reduces the linear response to better match the experimental results.  The non-linearities are also suppressed compared to the no mean field result, in good agreement with experiments. As it turns out this AFM mean field is also needed to account for the large negative $\theta_{CW}$ observed for this orientation in UPt$_3$. \\

\begin{figure}
\includegraphics[width=90mm]{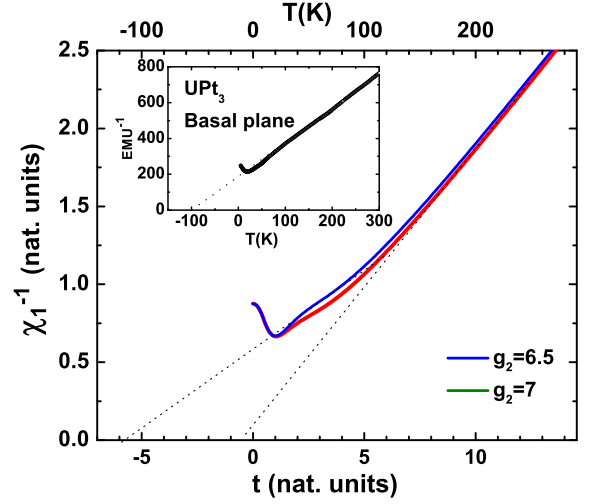}
\caption{\label{fig12}
Shows the high temperature behavior of $\chi_1$ for the parallel geometry. The two curves shown in the main figure correspond to two different values of $g_3$ as noted. A larger value shifts the curve to the left yielding a more negative $\theta_{CW}$. Removal of the FM mean field would shift the curve to even more to the left and would produce a $\theta_{CW}$ more negative than observed experimentally. The experimental results are shown in the inset.}
\end{figure}

\begin{figure}
 \includegraphics[width=93mm]{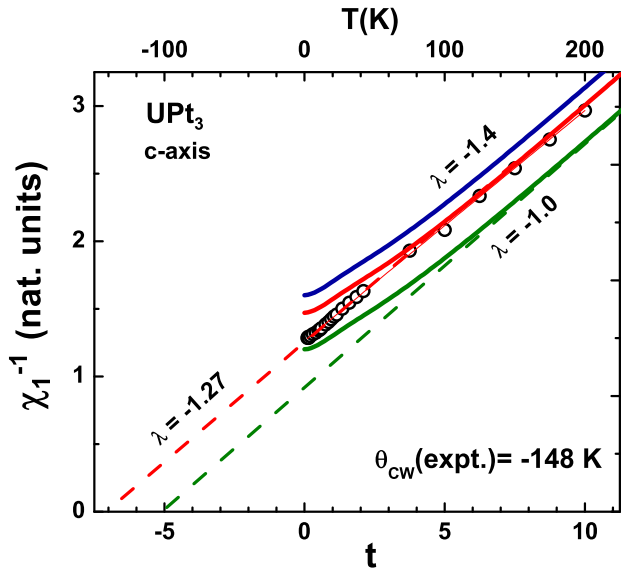}
\caption{\label{fig13}
Shows the high temperature behavior of $\chi_1$ for the perpendicular geometry. The three curves shown correspond to different $\lambda$s mentioned in the figure. The large negative $\theta_{CW}$ is obtained only with the $\lambda$ values shown. Remarkably, for this orientation the curves are insensitive to the values of the $\Delta$s chosen but the slopes of the curves depend on the g-factors.  However, these factors are 'fixed' by the values chosen with the analysis of the parallel geometry.}
\end{figure}

\textbf{Magnetic Properties - High Temperature Response:} The high temperature response in the SES+ model is very instructive. In the SES model  $\theta_{CW}$ was found to be equal to $\Delta$/3 for the parallel case. Thus based on this result we can surmise that if there were only two energy levels $\Delta_1$  and $\Delta_2$ and $\Delta_2 >> \Delta_1$ then essentially a single energy scale exists and that would provide a large Curie Weiss constant, $\Delta_2/3$. But to retain all the characteristic low temperature features such as the peak in $\chi_1$ and $\chi_3$ we need $\Delta_2$ to be $\approx k_BT_1$. Thus to obtain $ \theta_{CW} >> T_1$ in magnitude we need to add a third spin with a $\Delta_3  >>  \Delta_2$ to the Hamiltonian to effectively model the high temperature response. In fig. 12 we plot the inverse susceptibility for the parallel case with the same model parameters as above.  Clearly, it is possible to obtain a large negative $\theta_{CW}$ even though we have added a FM mean field. The value of $\theta_{CW}$ determined in most experiments is really that obtained from an intermediate temperature measurement - and in this range the model produces a "tilt' in the inverse susceptibility towards negative values. Thus there is a very smooth crossover in the intermediate T range - but it is remarkable that it appears as a straight line that yields a large negative Curie Weiss constant. \\

The high temperature response shown in fig. 13 for the perpendicular case is also instructive. The behavior, it turns out, is independent of the $\Delta$'s and while the slopes in fig.13 can be altered with the g-values these have already been chosen from an analysis of the data in the parallel geometry.  Thus in contrast to the parallel case the curves can only be shifted to the left with the choice of a negative $\lambda$. The insensitivity of the high temperature magnetic behavior to the $\Delta$'s in the perpendicular case is in line with the observed strong anisotropic uniaxial pressure dependence seen in many properties at low temperatures \cite{BoukhnyPhilMagLett1994}.\\

\begin{figure}
 \includegraphics[width=90mm]{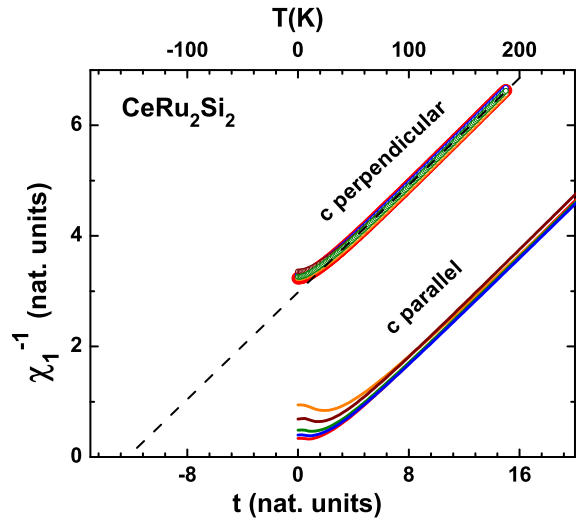}
\caption{\label{fig14}
Shows the high temperature behavior of $\chi_1$ for both the geometries in CeRu$_2$Si$_2$. The parameters used here are: $\Delta_1 =0.20, \Delta_2 = 1.95, \Delta_3 = 5.4, w = 0.55, g_1 = 0.84, g_2 = 1.55, g_3=1.7$ and are chosen by an analysis of the low temperature linear response.  Note that in contrast to UPt$_3$ there is a large negative intercept only for the perpendicular case. For the parallel direction the intercept is close to zero in agreement with experiments.} 
\end{figure}

\textbf{Heat Capacity:} Since the discovery of UPt$_3$ a significant point that has been raised is the presence of the $T^3ln(T)$ term in the heat capacity cited as the hallmark of  spin  fluctuations. This term since the early days of heavy fermion systems has been linked to the occurance of unconventional superconductivity in UPt$_3$ \cite{StewartPRL1984}. Such a term is also present in other heavy fermion metals such as UAl$_2$ which are not superconducting \cite{TrainorUAl2PRL1975}. In the following we demonstrate that $t^3ln(t)$ type behavior is also present in the current model. We explore the variation of this term on the parameters in the model and apply it to understand the pressure dependence of the heat capacity in UPt$_3$ in a later section. The heat capacity when spin  fluctuation contributions are present can be written as \\

$C =  \gamma T + \beta T^3 ln(T) + \epsilon T^3$. \\

Thus the quantity $(C/T -\gamma )/T^2$ captures just the logarithmic part separating out the linear Fermi liquid contribution and the phononic part of the heat capacity. Following the procedure adopted by Stewart \cite{Stewart1984} we show in fig. 15 the model heat capacity together with his experimental data.  The compliance between the two is remarkable. In this plot the vertical axis has been appropriately scaled to match the model with the experimental results. However, the scaling on the horizontal axis is fixed - it is derived from matching the peaks in $\chi_1$ as explained earlier.\\

\begin{figure}
 \includegraphics[width=90mm]{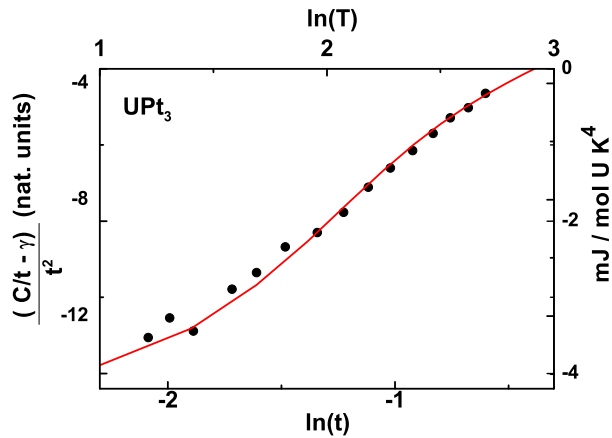}
\caption{\label{fig15}
Logarithmic term in the heat capacity and comparison to the experimental data of Stewart \cite{StewartPRL1984}.  The intercept on the vertical axis depends on $\epsilon$,the phononic part.  Since this contribution is absent in the model the values never reach zero on the left vertical axis. }
\end{figure}

\begin{figure}
 \includegraphics[width=90mm]{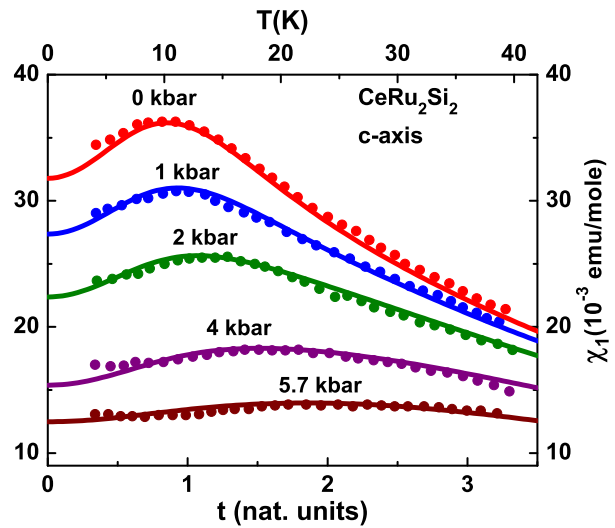}
\caption{\label{fig16}
The linear susceptibility in CeRu$_2$Si$_2$ at different pressures. The experimental points are from Voiron et. al. \cite{VoironCeRu2Si2HighPress1988}. The lines are fits from the present model. The parameters for zero pressure are: $\Delta_1 =0.20, \Delta_2 = 1.95, \Delta_3 = 5.4, w = 0.55, g_1 = 0.84, g_2 = 1.55$ and $g_3=1.7$. A plot of the variation of these parameters with pressure is provided in the supplementary section. The values of the g-factors are held fixed while the pressure is changed.}
\end{figure}

\textbf{Pressure and Substitution Dependence:} One of the hallmarks of heavy fermion materials is the strong pressure and composition dependence of the low temperature properties. In the case of UPt$_3$ and CeRu$_2$Si$_2$ these investigations were carried out soon after the discovery of metamagnetism \cite{BakkerFranseUPt3Press1992, VoironCeRu2Si2HighPress1988}.  Applying hydrostatic pressure the MM transition is shifted to higher fields in both systems with corresponding shifts in the position of the peak in the linear susceptibility to higher temperatures \cite{AokiCeRu2Si2PressJPSJ2011}. Although it would be useful to have the information in the context of the present model the uniaxial pressure dependence of MM has not yet been investigated in either system. Similarly, the pressure dependence of the third or higher order susceptibilities do not exist. However, a very nice study of the composition dependence of $\chi_3$ in CeRu$_2$Si$_2$ exists and it provides a good opportunity to test the current model by treating doping as being equivalent to pressure\cite{ParkVoironChi3CeRu2Si21994}. We begin however by considering the extensive measurements of the linear susceptibility performed by Voiron et al in CeRu$_2$Si$_2$ at various constant pressures. Through these measurements these researchers established several systematic trends: (a) the peak temperature T$_1$ shifts to higher values with increasing pressure, (b) there is a concomitant decrease in the zero temperature susceptibility, $\chi_1(0)$ and (c) the high temperature behavior is relatively unaffected. We can effectively model all of the observed trends by a simple addition of a uniform percentage change in all the parameters in going from low to high pressure (the precise values of the parameters at different pressures are shown in fig.S3 of the supplementary section).   

As mentioned earlier there are no measurements of the nonlinear susceptibility under pressure in any heavy fermion system but studies on $\chi_3$ on CeRu$_2$Si$_2$ with Y susbsitution are available. Yttrium being nonmagnetic has the simple effect of reducing the lattice parameter and has the same effect as pressure without introducing any additional changes in the electronic structure. The calculated temperature dependence of $\chi_3$ for the parameter values corresponding to the different pressures are shown in fig.17. Also shown in the inset of this figure are the measurements of Park et. al. \cite{ParkVoironChi3CeRu2Si21994}.  Here we made no particular attempt to match their results precisely - nevertheless the agreement is excellent. \\

Noteworthy in the model results as well as in the experiments are the crossing points - where $\chi_3$ has a common value at all pressures at one temperature (shown by the arrows in the figure). Such crossing points are a generic feature of strongly correlated systems \cite{VollhardtCrossingPRL1997} and in the context of heavy fermions can be seen in various other measurements such as heat capacity \cite{BrodaleCrossingPRL1986}. To our knowledge this is the first occurance of such a crossing in the nonlinear susceptibility.  \\

\begin{figure}
 \includegraphics[width=90mm]{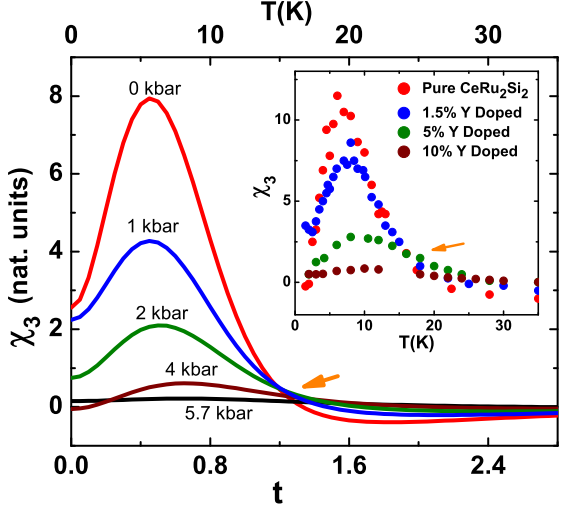}
\caption{\label{fig17}
The calculated nonlinear susceptibility for parameter values with the g-factors held fixed corresponding to the five different pressures measured by Voiron et al. \cite{VoironCeRu2Si2HighPress1988}.  The inset shows the experimental results of Park et al.  \cite{ParkVoironChi3CeRu2Si21994} which bear an amazing resemblance to the model results. Noteworthy in both the model results as well as in the experiments are the crossing points - where $chi_3$ has a common value at all pressures at one temperature - marked by the arrows. Such crossing points are a generic feature of strongly correlated systems.}
\end{figure}

Measurements of the pressure dependence of the magnetic properties in UPt$_3$ are somewhat less extensive.   Willis et al. \cite{WillisUPt3PressPRB1985} measured the linear susceptibility on polycrystalline samples and obtained a 17\% shift in $T_1$ from 17.6 K at zero pressure to 19.6 K at 4.7 kbar.  Bakker et al. studied the high field magnetism under pressure and found a linear increase in $H_c$ \cite{BakkerFranseUPt3Press1992}.  Given the linear correlation between $H_c$ and $T_1$ these experimental findings are qualitatively consistent.  For a more quantitative understanding we can turn to the model results.  Fig.18 shows the linear variation of h$_c$ with t$_1$ from the model together with the uniform percentage change in the $\Delta$s needed to cause the variation shown.  A key feature of the model is that while a change in the $\Delta$s markedly alters $t_1$ this temperature is completely insensitive to $\lambda$.  On the other hand $h_c$ is altered both by a change in the $\Delta$s and by $\lambda$.  This is shown in fig.19. where we plot a variation of $h_c$ under both scenarios - with $\lambda$ held fixed and secondly with the $\Delta$s helds fixed.  The experimental points from Bakker et al. are in better agreement with the first scenario.  These results are also consistent with the pressure dependence of the (zero field) heat capacity as discussed next.

The success of the present model extends beyond the magnetic properties as seen from the analysis of the heat capacity in UPt$_3$ at zero pressure where we successfully verified the presence of the logarithmic term in the model. Brodale et. al. \cite{BrodaleUPt3Press1986} performed very careful measurements of the heat capacity at high pressures and were able to use their measurements to quantify the various terms in the heat capacity equation  $C = \gamma T + \beta T^3 ln(T) + \epsilon T^3$. Employing the best fit values of the parameters, $\gamma$,  $\beta$ and $\epsilon$ provided by Brodale and co-workers we compute the expected experimental heat capacity at the three different pressures used and this is shown in fig.20, open circles. The solid lines in this figure show the model results with the parameters noted in the figure legend.  Since the model does not incorporte the phononic part appropriate values proportional to $T^2$ have been added to match the model results to the experiment on the high temperature side of the plot.  This addition has no effect on the low temperature side, which was exclusively used to determine the model parameters.  Thus the phononic part plays no role in influencing the values of the model parameters.  As can be seen from fig. 20 a pressure of 3.8 kbar requires a 5\% increase in the model parameters. Using this value in fig. 19 we find that the experimental values of $H_c$ are precisely what the model predicts (illustrated by the vertical dashed line and the solid red dot) .  \\

\begin{figure}
\includegraphics[width=80mm]{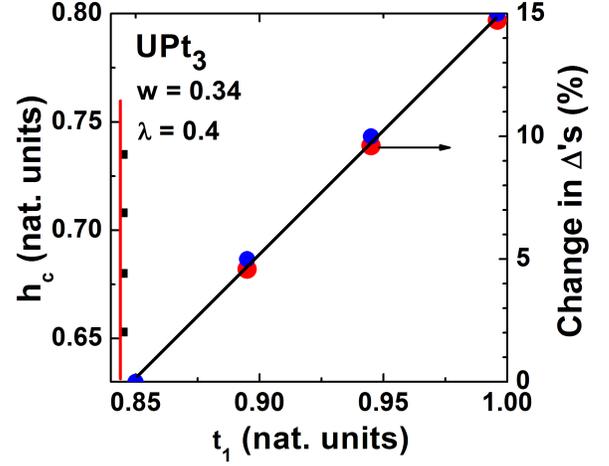}
\caption{\label{fig18}
The linear correlation of h$_c$ and t$_1$ in the model. Here we hold w and the g values fixed and consider both scenarios  - altering the $\Delta$s and holding $\lambda$ fixed or vice versa.  For the later scenario altering $\lambda$ is immaterial - the value of $t_1$ stays constant (vertical red line - black squares). The red cirlces indicate change in $\Delta$s needed for the shift in $t_1$ and the blue circles correspond to $h_c$ - left vertical axis. }
\end{figure}

\begin{figure}
\includegraphics[width=80mm]{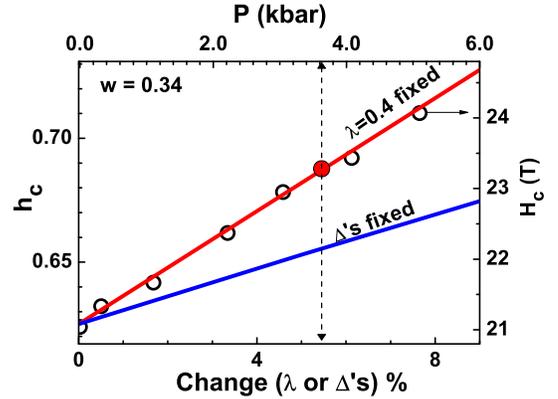}
\caption{\label{fig19}
Shows the dependence of model $h_c$ on the model parameters. The critical field is most sensitive to changes in $\Delta$ and increases relatively slowly with decrease in $\lambda$.  It is fairly independent of the hybridization parameter 'w'.  The g values are always held fixed.  Also shown in the figure are the experimental points from Bakker et al. \cite{BakkerFranseUPt3Press1992} - open circles.  The solid circle is the predicted value of $h_c$ from an analysis of the zero field heat capacity under pressure ( see fig. 20) }
\end{figure}

\begin{figure}
\includegraphics[width=80mm]{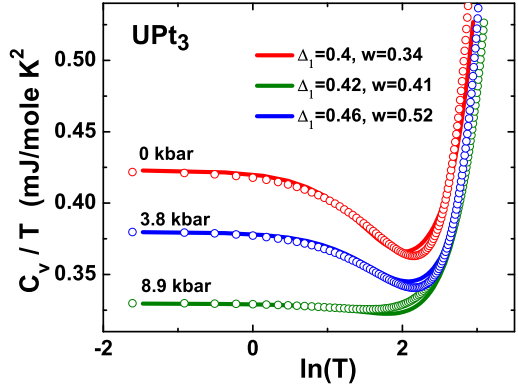}
\caption{\label{fig20}
The $C/T$ term in UPt$_3$ at different pressures and comparison with experiments. The experimental data is from Brodale et. al. The lines are fits from the present model.}
\end{figure}

\begin{figure}
\includegraphics[width=90mm]{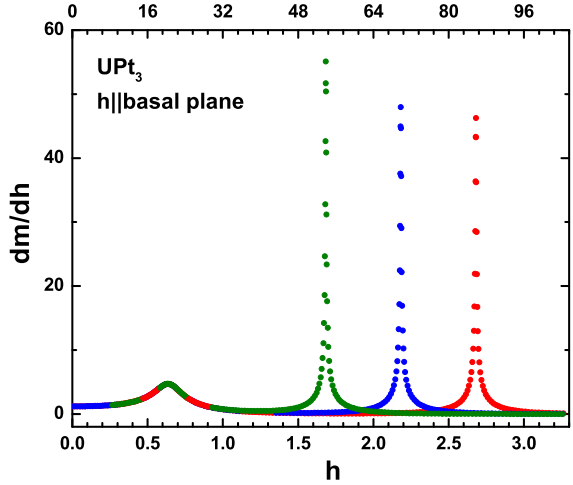}
\caption{\label{fig21}  
The differential susceptibility for fields upto 100 T in UPt$_3$. Note the large signature expected in the 60 to 100 T range. The precise value where this peak will occur depends on the value of $\Delta_3$ and g$_3$ chosen. We note that existing experiments on UPt$_3$ touch a maximum field of 45 T only and even at these high fields the magnetization has a strong upward slope and is not saturated. This is in agreement with our model results.}
\end{figure}

\begin{figure}
\includegraphics[width=90mm]{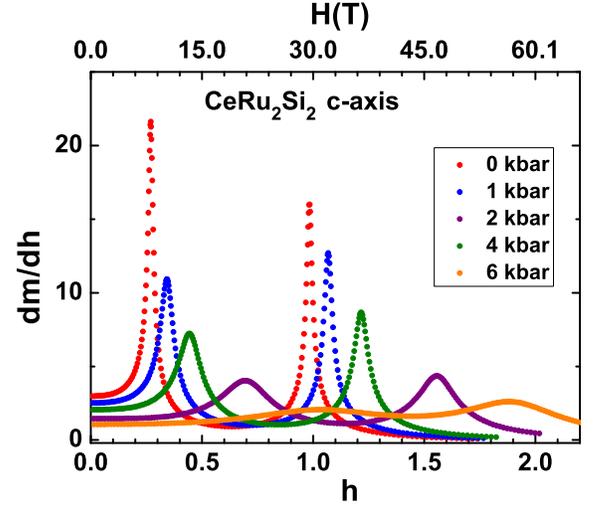}
\caption{\label{fig22} 
The differential susceptibility in high fields for CeRu$_2$Si$_2$. Note the large signature expected in the 30 to 70 T range. The precise value where this peak will occur depends on the value of $\Delta_3$ and g$_3$ chosen.}
\end{figure}

\textbf{Discussion:} It is clear from all of the above that there is a very strong agreement of the SES+ model with numerous existing experimental results not just in UPt$_3$ and CeRu$_2$Si$_2$ but in members of the family of alloys related to them. This level of agreement provides us with confidence to make several predictions. Obvious at first are further tests of scaling through nonlinear susceptibility measurements at higher pressures. But more significant are the qualitatively new predictions that arise from the SES+ model an example of which is provided in the two figures, fig.22 and 23.  These figures show the differential susceptibility in UPt$_3$ and CeRu$_2$Si$_2$ extended to currently available pulsed magnetic fields (such fields were non-existent at the time the two materials were discovered). In UPt$_3$ and CeRu$_2$Si$_2$ and typically in all other heavy fermions the moment is not saturated at the highest fields employed to date. This foretells that additional metamagnetic transitions can occur and indeed the careful analysis performed here reveals this. Apparent in the two figures are strong peaks in the differential susceptibility for both systems in the 40-100 tesla range with the one in UPt$_3$ being particularly prominent. There are other noteworthy features - in CeRu$_2$Si$_2$ the peak at the higher field grows at the expense of the lower one as the pressure is raised with the lower transition disappearing at 6 kbar.  Since somewhat smaller magnetic fields are involved in the case of CeRu$_2$Si$_2$ such a correlation could be easily checked. \\

Thus, it is remarkable that all of the analysis on data presented here and the predictions made come out of a very simple ''atomic'' model. This atomic picture is able to mimic the multiband structure common to heavy fermion materials \cite{SeyfarthMultiGap2008} and many other correlated systems such as the copper oxide superconductors \cite{MillisMultibandPRB2009} and the pnictides \cite{MultibandTesanovicEPL2008}. \\

 \begin{figure}
 \includegraphics[width=80mm]{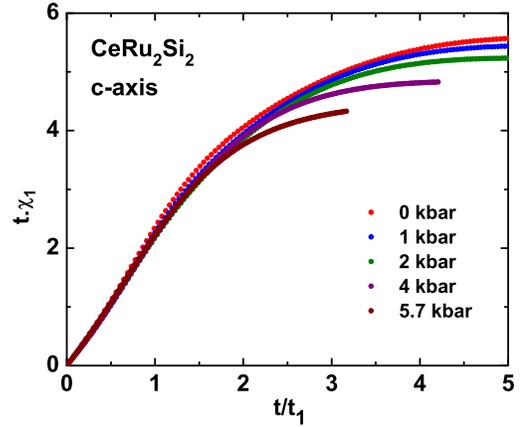}
\caption{\label{fig23} 
Scaling of the model results in analogy with the experimental plot, fig.5 of ref 37. The product $T\chi_1$ in the model is plotted as a function of the normalized temperature $T/T_1$ for the different pressures. The empirical plot is reproduced very well and the magnetic response is seen to collapse onto a single universal curve. This illustrates that a single energy type behavior is obtained despite the multiplicity of parameters. }
\end{figure}

\begin{figure}
 \includegraphics[width=90mm]{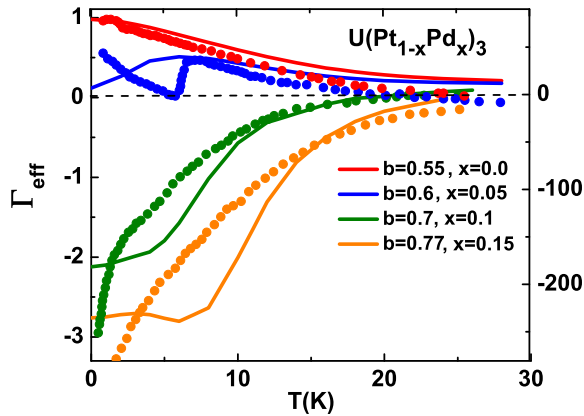}
\caption{\label{fig24} 
Shows the effective Gruneisen parameter, $\alpha^2/C_V$, obtained in the model for UPt$_3$ at different magnetic fields close to the critical field (left vetical axis).  The experimental curves (with corresponding values for $\Gamma_{eff}$ on the right vertical axis) are from de Visser et al \cite{deVisserFranseGruneiseninversionJMM1992} where doping was used as the control parameter to cross the quantum phase transition.  }
\end{figure}

At another level our approach is also similar in spirit to Landau's Fermi liquid theory where parametrization is employed to benchmark many body effects. In our model these parameters are the energy levels of the pseudo spins $\Delta_1, \Delta_2, \Delta_3$ and the effective g-factors, g$_1$, g$_2$ and g$_3$ together with the hybridization(Lorentzian) bandwidth 'w' and the anisotropic mean field $\lambda$.  We are able to accomplish a quantitative description of a multitude of experimentally measured quantities with varying just a few of these parameters. Many of them are fixed for any given material and only small changes to the remaining need to be considered to track the evolution of the experimental quantities with changes in pressure and composition for example. Despite such changes many "universal" correlations well known in the heavy fermion literature follow. Fig. 23 is an illustration of one such correlation where we plot the model results for the product $t\chi_1$ as a function of the normalized parameter t/t$_1$. The curves evaluated for parameters used to fit the experimental data for the five different pressures in fig. 16 collapse to reveal the universality. Yet another universal feature is revealed in fig. 24 which shows the 'effective Gruneisen parameter', $\alpha^2/C_V$, calculated in the model together with the experimental results from de Visser et. al. \cite{deVisserFranseGruneiseninversionJMM1992}. Here $\alpha$ is the volume expansion coefficient and C$_V$ is the constant volume heat capacity.  The thermal expansion in the model is proportional to the sum of the partial derivatives of the Helmholtz free energy with respect to the $\Delta$'s.  In the experimental result the effective Gruneisen parameter crosses across the zero line at low temperatures at the quantum phase transition, QPT, as the dopant concentration is changed. The control parameter to go through the QPT does not matter - in the model calculations it is the magnetic field with the parameters being the same as the ones used to fit (the low field) susceptibilities in UPt$_3$. The general agreement of the model results with experiments is again remarkable. \\

Other correlations that are well known in the heavy fermion literature such as the rapid increase in $\gamma$  as f-electron are spaced further apart and its eventual saturation\cite{MeisnerU2PtC2PRL1984} also come out in the model with remarkable fidelity. This can be understood in simple terms.  An increase in the f-electron spacing requires a negative pressure which is tantamount to reducing all three energy levels in the model. This has the tendency to create more excitations at low temperatures thus enhancing the heat capacity. Eventually however a point is reached where reducing the levels further compared to the hybridization parameter 'w' does very little to cause further enhancement. Thus the low temperature $\gamma$  saturates as indeed seen in the universal plot presented by Meisner et al.\\

With such overwhelming agreement with experiments nevertheless the question remains - what are we missing? Many of the heavy fermions order in some fashion with URu$_2$Si$_2$ for example exhibiting a unique form of hidden order. The two materials treated at length here though do not order magnetically but UPt$_3$ does superconduct at 0.5 K. Natural questions to ask are where do these differences come in, if at all possible, within the context of the SES+ model. These questions are currently being investigated and will be presented in forthcoming publications together with many other correlations that appear in the model. \\

\textbf{Acknowledgements:} The author wishes to express his gratitude to Pradeep Kumar and Vittorio Celli for their deep involvement and unwavering support during the course of this work. He is particularly thankful to Vittorio Celli for introducing him to the wonders of Maple the computation platform without which this work would have been impossible.   Also acknowledged are useful conversations with Peter Riseborough, Jim Smith, Greg Stewart and Kathy Levin.

 

\end{document}